\documentclass[aps,prl,twocolumn,superscriptaddress,groupedaddress]{revtex4}  

\usepackage{orcidlink}

\usepackage{graphicx} 
\usepackage{epstopdf} 
\usepackage{dcolumn} 
\usepackage{xcolor} 
\usepackage{amssymb} 
\usepackage{hyperref} 
\usepackage[utf8]{inputenc} 
\usepackage[T1]{fontenc}
\usepackage[english]{babel} 
\usepackage[displaymath, mathlines]{lineno} 
\usepackage{blindtext} 
\usepackage{amsmath}
\usepackage{siunitx}
\usepackage{xcolor}

\graphicspath{{figures/}} 

\newcommand {\darkh} {$e^{+}e^{-}\rightarrow A^\prime \, h^\prime$ with $A^\prime \rightarrow \mu^+\mu^-$ and $ h^\prime$ invisible}

\newcommand {\aprime}{$A^\prime$}
\newcommand {\hprime}{$h^\prime$}
\newcommand {\aprimemass}{$M_{A^\prime}$}
\newcommand {\hprimemass}{$M_{h^\prime}$}
\newcommand {\gevcc}{GeV/$c^2$}

\def\mumu       {\ensuremath{\mu^+\mu^-}\xspace}

\def\mumugamma       {\ensuremath{\mu\mu\gamma}\xspace}
\def\emu       {\ensuremath{e\mu}\xspace}
\def\babar{\mbox{\slshape B\kern-0.1em{\smaller A}\kern-0.1em
    B\kern-0.1em{\smaller A\kern-0.2em R}}}

\newcommand {\figsize} {0.95}

\begin{document}
\begin{flushleft}
Belle II Preprint 2022-001\\
KEK Preprint 2022-6
\end{flushleft}

\title{Search for a Dark Photon and an Invisible Dark Higgs Boson in $\mu^+\mu^-$ and Missing Energy Final States with the Belle II Experiment}

  \author{F.~Abudin{\'e}n\,\orcidlink{0000-0002-6737-3528}} 
  \author{I.~Adachi\,\orcidlink{0000-0003-2287-0173}} 
  \author{L.~Aggarwal\,\orcidlink{0000-0002-0909-7537}} 
  \author{H.~Aihara\,\orcidlink{0000-0002-1907-5964}} 
  \author{N.~Akopov\,\orcidlink{0000-0002-4425-2096}} 
  \author{A.~Aloisio\,\orcidlink{0000-0002-3883-6693}} 
  \author{N.~Anh~Ky\,\orcidlink{0000-0003-0471-197X}} 
  \author{D.~M.~Asner\,\orcidlink{0000-0002-1586-5790}} 
  \author{H.~Atmacan\,\orcidlink{0000-0003-2435-501X}} 
  \author{T.~Aushev\,\orcidlink{0000-0002-6347-7055}} 
% ...
\collaboration{The Belle II Collaboration}

\begin{abstract}
\vspace*{20pt}
The dark photon \aprime\ and the dark Higgs boson \hprime\ are hypothetical particles predicted in many dark sector models. 
We search for the simultaneous production of \aprime\ and \hprime\ in the dark Higgsstrahlung process \darkh\
in electron-positron collisions at a center-of-mass energy of 10.58~GeV in data collected by the Belle~II experiment in 2019. With an integrated luminosity of 8.34~fb$^{-1}$, we observe no evidence for signal.
We obtain exclusion limits at 90\% Bayesian credibility in the range of 1.7--5.0~fb on the cross section and in the range of $1.7 \times10^{-8}$--$200 \times10^{-8}$ on the effective coupling $\varepsilon^2 \times \alpha_D$ for the \aprime\ mass in the range of 4.0~\gevcc\ $< M_{A^\prime}< 9.7$~\gevcc\ and for the \hprime\ mass $M_{h^\prime} < M_{A^\prime}$, where $\varepsilon$ is the mixing strength between the standard model and the dark photon and $\alpha_D$ is the coupling of the dark photon to the dark Higgs boson.
Our limits are the  first in this mass range.   \newline
\vspace*{10pt}
\begin{center}  
This paper is dedicated to the memory of Cate MacQueen.
\end{center}    
\end{abstract} 
\maketitle

Several astrophysical observations in the last decades suggest the existence of a large 
quantity of dark matter in the universe coupled with ordinary  matter, at least through gravitational interactions. 
In recent years, theoretical models (commonly called dark sector models) with light particles mediating new interactions have gained considerable attention as solutions to  the long-standing problem of  reproducing the observed relic density  of dark matter.
A  well-motivated  model predicts the existence of an additional massive vector gauge boson, a dark photon~\aprime, coupled to the standard model (SM)  only through its kinetic mixing $\varepsilon$ with the hypercharge field~\cite{POSPELOV200853, PhysRevD.79.015014, ALVES2010323, POSPELOV2009391, PhysRevLett.92.031303, CIRELLI20091, March_Russell_2008, PhysRevD.80.123518, Cholis_2009, Arkani_Hamed_2008}.
In this model, dark matter particles can annihilate to SM particles, and vice versa, through the exchange of dark photons: this process contributes to the relic density of dark matter so as to match the observed value.
The mass of the \aprime\ can arise from spontaneous symmetry breaking, which introduces a new scalar particle: a dark Higgs boson \hprime~\cite{PhysRevD.79.115008}. 
Several searches for the \aprime\ have set upper limits on $\varepsilon$~\cite{PhysRevLett.106.251802, PhysRevLett.112.221802, PhysRevLett.107.191804, 2013187, 2014265, BABUSCI2013111, 2014459, 2016356, ANASTASI2015633, 2018336, PhysRevLett.113.201801, 2017252, PhysRevLett.124.041801}; they depend on the mass of the dark photon and are typically of the order $\varepsilon^2 \leq 5 \times 10^{-7}$ for masses below 10.6~\gevcc. 

In this Letter, we search for the so-called dark Higgsstrahlung process $e^{+}e^{-} \rightarrow A^\prime h^\prime$ using data collected by the Belle~II experiment at the SuperKEKB collider.
We consider the minimal secluded model of Ref.~\cite{PhysRevD.79.115008}, in which either the dark Higgs boson does not 
mix with the SM Higgs boson or its mixing can be neglected and any additional particles (in particular dark matter candidates) are heavier than both \aprime\ and \hprime.
The cross section for dark Higgsstrahlung is proportional to $\varepsilon^2 \times \alpha_{D}$, where $\alpha_{D}$ is the coupling constant 
of the secluded model~\cite{PhysRevD.79.115008}.
Two  scenarios exist, differentiated by the hierarchy of the masses \aprimemass\ and \hprimemass. 
If $M_{h^\prime} > M_{A^\prime}$, then the \hprime\ decays dominantly to an \aprime$A^{\prime(*)}$ pair (where $A^{\prime(*)}$ can be virtual), which is a  final state searched for by the BaBar~\cite{PhysRevLett.108.211801} and Belle~\cite{PhysRevLett.114.211801} experiments.
If $M_{h^\prime} < M_{A^\prime}$, then the \hprime\ is long lived  and invisible because it does not interact with the detector material.
We focus on the latter scenario, which was previously investigated by KLOE~\cite{2015365} but in a much smaller mass range than is accessible to Belle~II.
Our search is limited to the decay of the \aprime\ into a muon pair. 

The dark Higgsstrahlung process studied here produces a pair of oppositely charged muons with a mass $M_{\mu\mu}$ distributed around \aprimemass\ and a recoil against them with a mass $M_{\text{recoil}}$ that,
in the absence of radiated photons, is distributed around \hprimemass.
In the following, recoil quantities are computed against the dimuon system.
The accessible search region is the mass plane $M_{\text{recoil}}$--$M_{\mu\mu}$: it has a triangular shape, limited by the conditions $M_{\text{recoil}}<M_{\mu\mu}$ (corresponding to $M_{h^\prime} < M_{A^\prime}$) 
and $M_{\text{recoil}}+M_{\mu\mu} \leq \sqrt{s}/c^2$, resulting from energy conservation. 
Our analysis uses events with exactly two tracks, identified as muons, and negligible extra energy.
The backgrounds are SM processes that produce final states with two tracks identified as muons and missing energy. 
These are dominantly $e^+e^- \rightarrow \mu^+\mu^-(\gamma)$ with typically one or more photons  undetected due to inefficiency or limited acceptance; $e^+e^- \rightarrow \tau^+\tau^-(\gamma)$ with both $\tau$ leptons decaying to muons and neutrinos; and $e^+e^- \rightarrow e^+e^-\mu^+\mu^-$ with electrons outside acceptance. We search for the signal as a narrow enhancement in the two-dimensional $M^2_{\text{recoil}}$--$M^2_{\mu\mu}$  distribution.  
We scan the search region for local excesses above the expected background with a counting technique that uses a set of two-dimensional $M^2_{\text{recoil}}$--$M^2_{\mu\mu}$ overlapping windows. 
Event selection is optimized using simulated events prior to examining data.

The Belle~II detector~\cite{Abe:2010sj} operates at the SuperKEKB electron-positron $e^+e^-$ collider~\cite{superkekb}, located at  KEK in Tsukuba, Japan. 
The beam energies are 7~GeV for $e^-$ and 4~GeV for $e^+$, resulting in a boost of $\beta\gamma = 0.28$ of the center-of-mass (c.m.) frame relative to the laboratory frame.
Data used in this analysis were collected in 2019 at a c.m.~energy $\sqrt{s}$ corresponding to the mass of the $\Upsilon$(4S) resonance, for an integrated luminosity of $8.34 \pm 0.08$~fb$^{-1}$ (see Ref.~\cite{lumi} for the description of the luminosity measurement technique).

The Belle~II detector consists of several subdetectors arranged around the beam pipe in a cylindrical structure.
Subdetectors relevant for this analysis are briefly described here in the order of innermost out; a description of the full detector is given in Refs.~\cite{Abe:2010sj, ref:b2tip}.
The innermost subdetector is the vertex detector, which consists of two inner layers of silicon pixels and four outer layers of silicon strips.
The second pixel layer was only partially installed for the data sample we analyze, covering only one sixth of the azimuthal angle.
The main tracking subdetector is a large helium-based small cell drift chamber (CDC).
The relative charged-particle transverse momentum resolution is typically 0.1\% $p_T$ $\oplus$ 0.3\%, with $p_T$ expressed in GeV/$c$.  
An electromagnetic calorimeter (ECL) consists of a barrel and two end caps made of CsI(Tl) crystals.
A superconducting solenoid, situated outside of the calorimeter, provides a 1.5~T magnetic field.
A $K^0_L$ and muon subdetector (KLM) is made of iron plates, which serve as a magnetic flux-return yoke, alternated with resistive-plate chambers and plastic scintillators in the barrel as well as with plastic scintillators only in the end caps.
The longitudinal direction, the transverse plane and the polar angle $\theta$ are defined with respect to the detector's solenoidal axis in the direction of the electron beam. 
In the following, quantities are defined in the laboratory frame unless specified otherwise.

The identification of muons uses criteria that rely mostly on  track penetration in the KLM for momenta larger than 0.7~GeV/$c$ and on information from the CDC and ECL otherwise.
Electrons are identified mostly by comparing measured  momenta in the CDC with energies of the associated ECL clusters.
Photons are identified as ECL clusters with energies greater than 100~MeV that are not associated with tracks.
Details of the particle reconstruction and identification algorithms are in Refs.~\cite{ref:b2tip, tracking, pid}.

Simulated signal events are generated using \texttt{MadGraph5}~\cite{Alwall2014} with and without initial-state radiation (ISR) for $A^{\prime}$ masses ranging from 0.21 to 10.45~\gevcc\ and \hprime\ masses ranging from 0.01~\gevcc\ to the minimum of ($M_{A^\prime}$, $\sqrt{s}/c^2-M_{A^\prime}$) in variable steps that follow the mass resolutions.
The resolution on \aprimemass\  ranges between 3 and 50~MeV/$c^2$, with an average value of 25~MeV/$c^2$. The resolution on \hprimemass\ ranges between 30 and 900~MeV/$c^2$, with an asymmetric distribution peaking at 100~MeV/$c^2$.
We generate samples of 10~000 events  for each of the 9003 pairs of \aprimemass\ and \hprimemass.

The considered background processes are simulated using the specified generators:
$\mu^+\mu^-(\gamma)$  with \texttt{KKMC}~\cite{ref:kkmc};
$\tau^+\tau^-(\gamma)$ with \texttt{KKMC} interfaced with \texttt{TAUOLA}~\cite{ref:tauola};
$ e^+e^-\mu^+\mu^-$ and $ e^+e^-e^+e^-$ with \texttt{AAFH}~\cite{ref:fourlepton};
$e^+e^-\pi^+\pi^-$ with \texttt{TREPS}~\cite{uehara2013treps};
$\pi^+\pi^-(\gamma)$ with \texttt{PHOKHARA}~\cite{ref:phokhara};
$e^+e^-(\gamma)$ with \texttt{BabaYaga@NLO}~\cite{ref:babayaga};
$b\bar b$ with \texttt{EvtGen}~\cite{evtgen};
$q\bar q$ ($q=u,d,s,c$) with \texttt{KKMC} interfaced with \texttt{PYTHIA8}~\cite{pythia8} and \texttt{EvtGen};
$J/\psi\gamma$, $\psi(2S)\gamma$ with $J/\psi, \psi(2S) \rightarrow \mu^+\mu^-$ with \texttt{PHOKHARA}; and $\mu^+\mu^- \nu \bar\nu$ with \texttt{KoralW}~\cite{koralw}. Only  $\mu^+\mu^-(\gamma)$, $\tau^+\tau^-(\gamma)$, and $ e^+e^-\mu^+\mu^-$ actually contribute to the background, with the others being negligible. 

The detector geometry and interactions of final-state particles with detector materials are simulated using \texttt{\textsc{Geant4}}~\cite{ref:geant4} and the Belle~II Analysis Software Framework (\texttt{basf2})~\cite{basf2}. Both real data and simulated events are reconstructed and analyzed using \texttt{basf2}.

The search uses an online event selection (trigger) that requires at least one pair of tracks in a restricted polar angle acceptance,   $\theta\in[37,120]^\circ$, with an azimuthal opening angle $\Delta \phi$ larger than $90^\circ$ and rejects events consistent with Bhabha scattering through a dedicated veto based on the pattern of energy depositions in the ECL. 
The efficiency of this trigger for $\Delta \phi > 90^\circ$ is 89\%, measured in $\mu^+\mu^-(\gamma)$  events.
The trigger requirement on the opening angle leads to large signal inefficiencies for $M_{A^\prime} < 4$~\gevcc. 

To suppress misreconstructed and beam-induced background tracks, we require that the transverse and longitudinal projections of their distances of closest approach to the  interaction point be smaller than 0.5 and 2.0~cm, respectively.
We require that events have exactly two oppositely charged  particles, identified as muons, with polar angles  $\theta\in[37,120]^\circ$ and with $\Delta \phi$ larger than $90^\circ$ to match the trigger requirements. 
In addition, the energy of each ECL cluster associated with a muon track must be below 1.5~GeV to suppress background events from muon pairs, with final-state radiation and unresolved photons close to the muons. 
We require that the recoil momentum point into the ECL barrel, $\theta\in[32,125]^\circ$, to exclude regions where photons from radiative backgrounds can escape undetected.
This selection also increases the signal-to-background ratio by suppressing  $\mu^+\mu^-(\gamma)$ and $e^+e^-\mu^+\mu^-$ processes that, unlike the signal, have recoil momenta dominantly in the forward direction.
To reduce radiative muon-pair backgrounds, we require that the total energy of all photons be less than 0.4~GeV and no photon be within $15^\circ$ of the recoil momentum.
To suppress background events from $\mu^+\mu^-(\gamma)$ and $e^+e^-\mu^+\mu^-$ processes, we require that the transverse c.m. frame momentum of the muon pair be greater than 0.1~GeV/$c$. 

We count events in 9003 partially overlapping regions (search windows) of the two-dimensional space of squared dimuon and recoil masses, which span the $M_{\text{recoil}}$--$M_{\mu\mu}$ search plane. We search for signal by comparing the observed yields with expectations from known backgrounds.
The squared dimuon and recoil masses are negatively correlated in signal events and, to a lesser extent, in background events, with the correlation varying across the plane.
Initial-state radiation partially spoils this correlation because it affects the $M_{\text{recoil}}$ distribution only.
The search window boundaries are chosen as ellipses in the $M^2_{\text{recoil}}$--$M^2_{\mu\mu}$ plane that take the local correlation into account.
Each window is centered at one of the values of ($M^2_{A^\prime}$, $M^2_{h^\prime}$) used to produce a simulated dataset.
We fit a sum of two two-dimensional Gaussian distributions that share a common mean and correlation to simulated data without ISR.
The elliptical search window boundaries correspond to the two-dimensional two-standard-deviation contours resulting from the fit.
Search windows partially overlap to maximize signal efficiency, with an overlap in area typically around 75\%. 
The average fraction of signal events retained in a search window is 71\%, with variations due to the different effects induced by ISR depending on \aprimemass\ and \hprimemass. 
Accounting for correlation in defining the window increases the signal-to-background ratio by a factor of three to five.

A final selection is based on the helicity angle $\eta$, defined as the angle in the dimuon rest frame between the momentum direction of the c.m. system and the momentum direction of the $\mu^-$.
The $\eta$ distribution for the signal is that of a massive vector particle decaying into two fermions.
For an unpolarized \aprime,  the distribution of $C_\eta = \vert \cos \eta \, \vert$ is uniform. 
In background events, $C_\eta$ peaks at one because the muons come either from independent decays, as in $\tau^+\tau^-(\gamma)$, or from physics processes of a different nature, as in $\mu^+\mu^-(\gamma)$ and $e^+e^-\mu^+\mu^-$.
We exclude high values of $C_\eta$, typically larger than 0.9,  by maximizing the figure of merit of Ref.~\cite{Punzi:2003bu} in each search window.

The resulting signal efficiency depends on the masses of the \aprime\ and \hprime\ and ranges between 10 and 25\% for $M_{A^\prime} > 4$~\gevcc.
The efficiency drops considerably for $M_{A^\prime} < 4$~\gevcc\ and becomes too low to allow  analysis for $M_{A^\prime} < 1.65$~\gevcc.
We restrict our analysis to the range $M_{A^\prime} > 1.65$~\gevcc.
Backgrounds are typically reduced by factors of 10--1000.
The surviving background events come 78.5\% from $\mu^+\mu^-(\gamma)$, 18.5\% from $\tau^+\tau^-(\gamma)$, and 3\% from $e^+e^-\mu^+\mu^-$: they dominantly populate the regions of the search plane close to the kinematical limit $M_{\text{recoil}} + M_{\mu\mu} \approx \sqrt{s}/c^2$, where the $C_\eta$ selection is less effective and   $\mu^+\mu^-(\gamma)$ contributes more, leaving most of the mass plane almost background free.

The distribution of the observed event counts in the search windows is shown in Fig.~\ref{fig:data}.
Because of overlap, events can be counted  in multiple search windows: this correlation creates groups of isolated, sparsely populated windows. 
A total of 28~985 events pass all the selection criteria, which is in agreement with the expectation from simulation, 28~486 $\pm$ 331.
The sum of the event counts inside all mass windows is 78~740.
On average, each event is contained in 2.7 search windows.

We search for an excess of signal above the expected background independently in the 9003 search windows. 
Event counts $N$ in a search window are interpreted according to the relation $N = \epsilon_{\text{sig}} \times L \times \sigma  + B$, where $\sigma$ is the cross section for the dark Higgsstrahlung process \darkh, $L = \int \mathcal{L} \, dt$ is the integrated luminosity, and $\epsilon_{\text{sig}}$ and $B$ are the signal efficiency and the expected background inside the window.
Both $\epsilon_{\text{sig}}$ and $B$ are determined from  simulation and are subject to systematic uncertainties.

\begin{figure}[htb] 
  \centering
    \includegraphics[width=\figsize\linewidth]{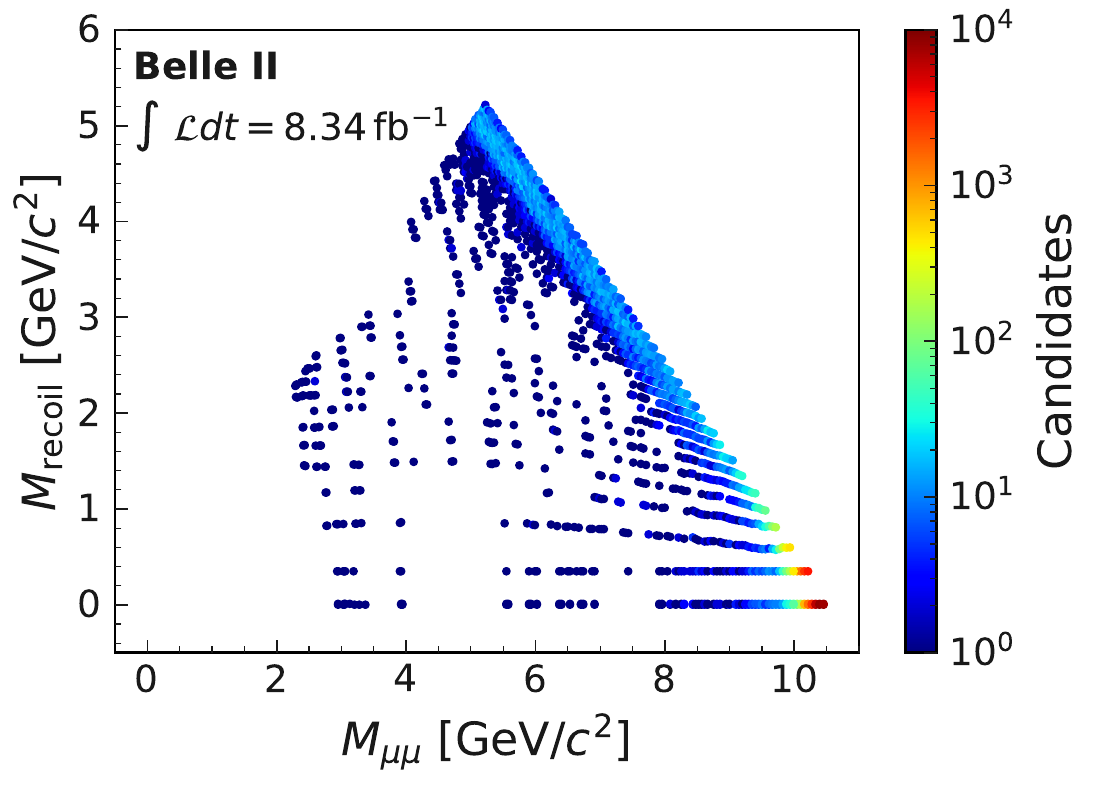}
    \caption{Observed event counts inside the search windows after all selection criteria. Points correspond to search window centers. Search window boundaries are not shown.} 
    \label{fig:data}
\end{figure}

Several sources of systematic uncertainties affecting the signal efficiency and the background estimate are taken into account.
They are studied by comparing data and simulation on two control samples that emulate  the two main background processes $\mu^+\mu^-(\gamma)$ and $\tau^+\tau^-(\gamma)$ and on the search \mumu\ sample. 
The  \mumugamma control sample contains events that pass all selection criteria except for the veto on the presence of photons, which is replaced  by a requirement that a photon with energy greater than 1~GeV be reconstructed in the barrel of the ECL. This sample is dominantly composed of $\mu^+\mu^-(\gamma)$ events.
The \emu control sample contains events passing all the selection criteria but with an identified electron replacing an identified muon. This sample is almost entirely composed of $\tau^+\tau^-(\gamma)$ events. 
We split the mass plane into six non-overlapping macroregions, with each dominantly populated by a single background source.
Events are counted in the data and simulation for each macroregion and their discrepancies used to evaluate systematic uncertainties. 
When using the search \mumu\ sample, a region  ten times larger than  the search window under study  is excluded from the counting to avoid that the presence of a signal may bias the result.

Uncertainties affecting the background  due to the trigger, luminosity, tracking efficiency, muon identification, cross sections, and the selection criteria are collectively evaluated through the macroregion studies before applying the $C_\eta$ selection.
Over most of the mass plane, discrepancies between the data and simulation, for both the control and the search \mumu\ samples, are of the order of 2\%, which are treated as relative systematic uncertainties. 
In a small region, where $M_{A^{\prime}} > 9$~\gevcc, the search \mumu\ sample data yields are 9.1\% lower than the simulation yields.
In this region, there are also discrepancies between the data and simulation in the shapes of mass distributions that lead to an additional relative systematic uncertainty of 9.3\%: we sum it quadratically with the 9.1\% normalization uncertainty.  
We assume the 2\% and 9.1\% uncertainties also hold  for the signal efficiency below and above $M_{A^{\prime}} = 9$~\gevcc, respectively.

Uncertainties affecting the background due to the $C_\eta$ selection are evaluated by comparing the data and simulation.
The numbers of observed and expected events  for both the control and the search \mumu\ samples agree within 1\%, which we use as a relative systematic uncertainty due to this source.
We evaluate the contribution due to this effect on the signal efficiency to be negligible.

We also include systematic uncertainties due to discrepancies in the dimuon and recoil mass resolutions in data and simulation.
A modified \mumugamma control sample is used to check the mass distributions in the region of the $J/\psi$ resonance: the requirement on the opening angle $\Delta \phi$ is released and an ECL-only trigger is used, which is allowed by the presence of the photon. 
This trigger requires  that the total energy deposition in the barrel and in the forward end cap exceed 1 GeV. 
The search \mumu\ sample is also used to check the mass distributions in the region where the dimuon mass is close to $\sqrt{s}/c^2$.
We find differences between the data and simulation in the dimuon and recoil mass resolutions of no more than $10\%$.
Their effects on the relative signal efficiency range between 1 and 5\%, depending on the masses, with an average of 2.4\%.

We evaluate a systematic uncertainty on the signal efficiency due to \aprimemass\ and \hprimemass\ not coinciding with a window center by recalculating the efficiency with the two mass values randomly varied to points near the window center. The signal efficiency varies 2\% on average, and no more than 5\%, which we assign as a relative systematic uncertainty.

Finally, a relative systematic uncertainty of 4\% on the theoretical prediction of the \aprime\ decay branching fraction to muons is used when interpreting results in terms of the coupling product $\varepsilon^2\times \alpha_{\text{D}}$.
This uncertainty comes dominantly from uncertainties on the measured ratio of cross sections for the production of hadrons or muons in $e^+e^-$ collisions, which enter in the \aprime\ width theoretical calculation~\cite{PhysRevD.79.115008}.

The average total relative systematic uncertainties are 2.2 and 5.4\% on the background and signal efficiencies, respectively.
They rise up to 12.7 and 11.3\% in the region $M_{A^{\prime}} > 9$~\gevcc.

We search for excesses in data in each window separately with both a Bayesian technique based on Bayes factors~\cite{bayes_factors} and a frequentist technique based on significances from one-sided Gaussian integral transformation of $p$~values. 
Background expectations and signal efficiencies are assumed from the simulation. Systematic uncertainties are taken into account as correlated Gaussian smearings of background expectations and signal efficiencies, with  widths equal to the estimated uncertainties. 
We choose thresholds of 80 for the Bayes factor and of  3.5$\sigma$ for the significance before inspecting the data: they are larger than normally used, because we expect a relevant look-elsewhere effect~\cite{Gross:2010qma, Vitells:2011da} due to the high number of search windows. 
We find only one case of a local significance above the threshold, 3.7$\sigma$, which also corresponds to the highest Bayes factor of 45.6.
It is in the search window centered at $M_{\mu\mu} = 5.44$~GeV/c$^2$ and $M_{\text{recoil}} = 3.18$~GeV/c$^2$.
Taking into account the look-elsewhere effect, this excess has a global significance below $1 \sigma$, showing no evidence for signal.

\begin{figure}[htb] 
  \centering
    \includegraphics[width=\figsize\linewidth]{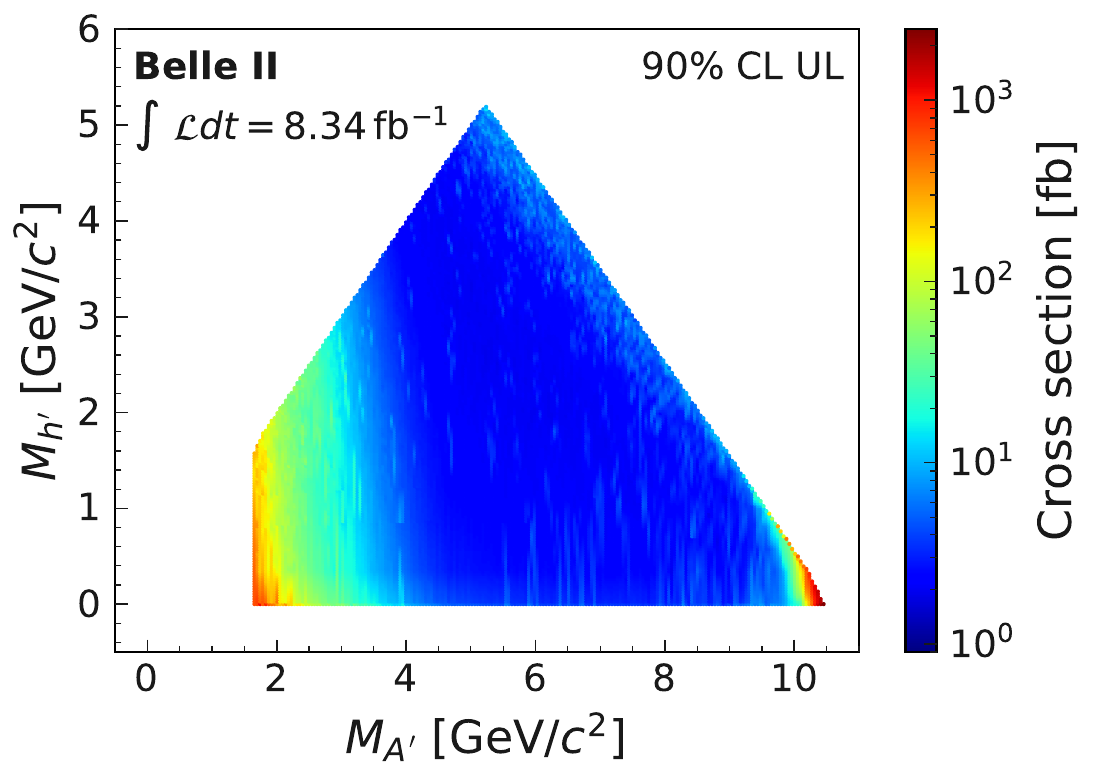}
    \caption{Observed 90\% CL upper limit on the cross section of \darkh\  as a function of the \aprime\ and \hprime\ masses. Values are computed at search window centers and then interpolated to points of the search plane.} 
    \label{fig:UL_CS}
\end{figure}

We compute Upper Limits (UL) at a 90\% Bayesian credibility level (CL)  on the cross section for the dark Higgsstrahlung process \darkh\ as a function of \aprimemass\ and \hprimemass\ using the Bayesian Analysis Toolkit software package~\cite{bat_software}.
We assume uniform priors for all positive values of the cross section, Poissonian likelihoods for the number of observed and simulated events, and Gaussian smearing to model  systematic uncertainties, accounting for their correlations.
The result is shown in Fig.~\ref{fig:UL_CS}.

\begin{figure}[!htb] 
  \centering
    \includegraphics[width=\figsize\linewidth]{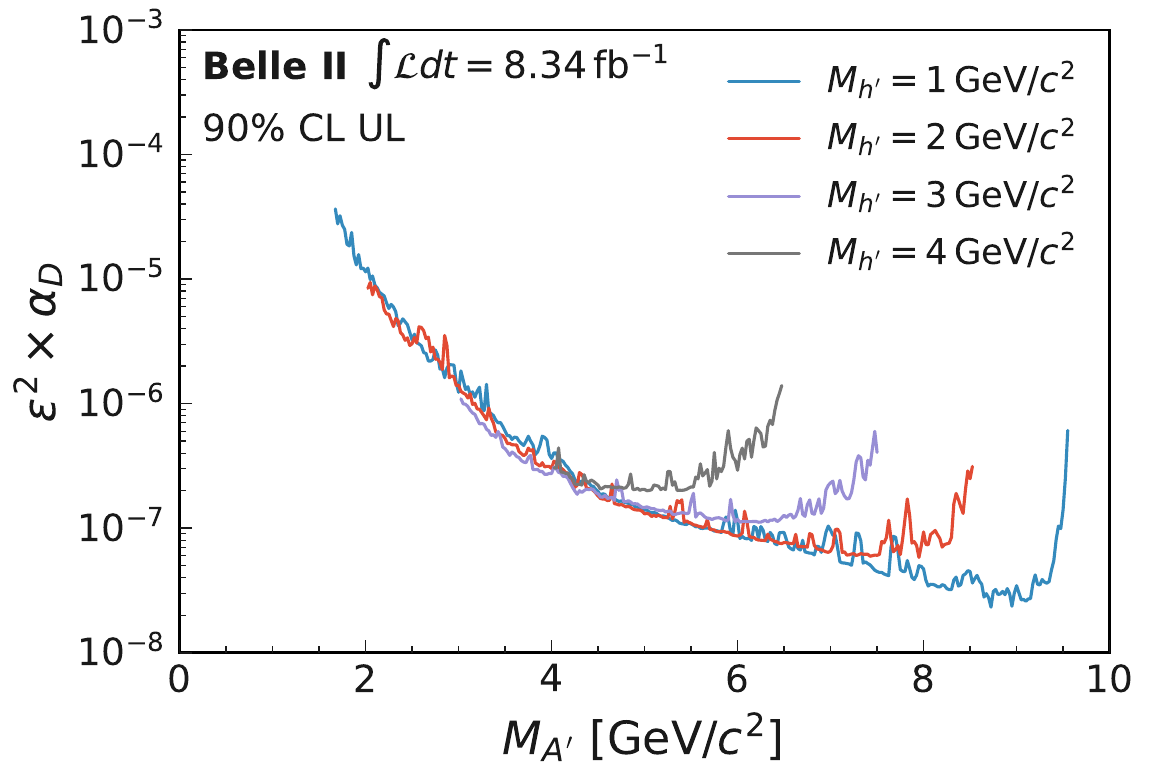}
    \includegraphics[width=\figsize\linewidth]{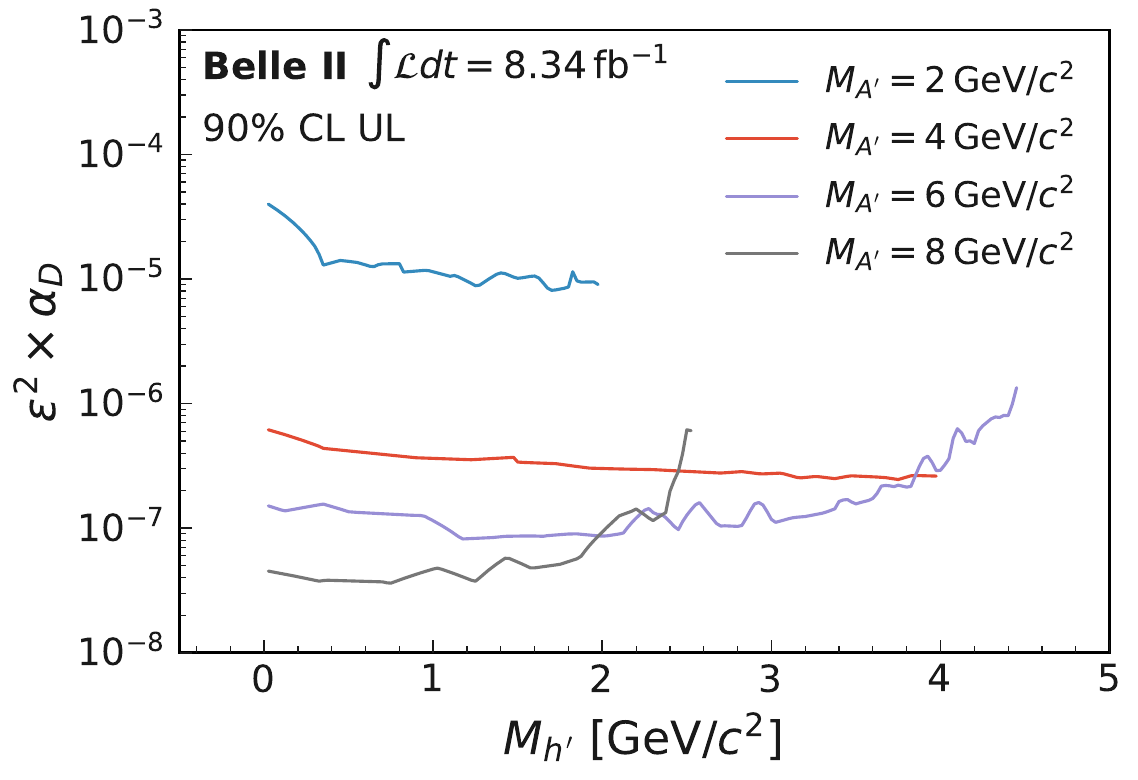}
    \caption{Observed 90\% CL upper limits on $\varepsilon^2 \times \alpha_{D}$ ({\it top}) as functions of \aprimemass\ for four values of \hprimemass\  and ({\it bottom}) as  functions of \hprimemass\ for four values of \aprimemass.} 
    \label{fig:UL_couplings}
\end{figure}

We translate the cross section result into 90\% CL upper limits on  $\varepsilon^2 \times \alpha_{D}$.
These limits are shown in Fig.~\ref{fig:UL_couplings} as functions of \aprimemass\ for four different values of \hprimemass\ and as functions of \hprimemass\ for four different values of \aprimemass.\

Our results are dominated by their statistical uncertainties.
In most of the search plane, systematic uncertainties degrade the upper limits by less than 1\%.
Only in the small region where $M_{A^\prime} > 9$~\gevcc\ are systematic uncertainties significant, worsening the upper limits by 25\%.
We test for prior dependence of the results by using logarithmic priors for the cross section and find differences smaller than 3\%. 
Additional plots and detailed numerical results are provided in the Supplemental Material~\cite{supplemental}.

In summary, we search for the dark Higgsstrahlung process \darkh\ in a data sample of electron-positron collisions at 10.58~GeV collected by Belle~II at SuperKEKB in 2019, corresponding to an integrated luminosity of 8.34~fb$^{-1}$.
We find no significant excess above the expected background and set upper limits on the cross section and coupling $\varepsilon^2 \times \alpha_{D}$ for \aprimemass\ between 1.65 and 10.51~\gevcc\ and $M_{h^\prime} < M_{A^\prime}$. 
Our limits are the first in this mass range. The excluded region is much larger than that previously covered by other experiments~\cite{2015365}.
Our 90\% CL upper limits range between 1.7 and 5.0~fb for the cross section and between $1.7 \times 10^{-8}$ and $200 \times 10^{-8}$ for the coupling for 4.0~\gevcc\ $< M_{A^\prime} < 9.7$~\gevcc\ and $M_{h^\prime} < M_{A^\prime}$.
For specific values of $\alpha_{D}$ and assuming the existence of a light invisible dark Higgs, our results can be interpreted as upper limits on $\varepsilon^2$ and compared with limits obtained by other experiments. With $\alpha_{D} = 1$,  our constraints would improve on previous searches~\cite{PhysRevLett.113.201801} across almost the full mass range.
For $\alpha_{D} = 0.1$, this conclusion would still hold in a substantial part of the mass range. 
These results can be interpreted in a wider class of models compared to that of Ref.~\cite{PhysRevD.79.115008}, for example those with a long-lived invisible \hprime\ that mixes with the SM Higgs boson~\cite{Darme:2017glc, Duerr:2020muu}.


This work, based on data collected using the Belle II detector, which was built and commissioned prior to March 2019, was supported by
Science Committee of the Republic of Armenia Grant No.~20TTCG-1C010;
Australian Research Council and research Grants
No.~DE220100462,
No.~DP180102629,
No.~DP170102389,
No.~DP170102204,
No.~DP150103061,
No.~FT130100303,
No.~FT130100018,
and
No.~FT120100745;
Austrian Federal Ministry of Education, Science and Research,
Austrian Science Fund
No.~P~31361-N36
and
No.~J4625-N,
and
Horizon 2020 ERC Starting Grant No.~947006 ``InterLeptons'';
Natural Sciences and Engineering Research Council of Canada, Compute Canada and CANARIE;
Chinese Academy of Sciences and research Grant No.~QYZDJ-SSW-SLH011,
National Natural Science Foundation of China and research Grants
No.~11521505,
No.~11575017,
No.~11675166,
No.~11761141009,
No.~11705209,
and
No.~11975076,
LiaoNing Revitalization Talents Program under Contract No.~XLYC1807135,
Shanghai Municipal Science and Technology Committee under Contract No.~19ZR1403000,
Shanghai Pujiang Program under Grant No.~18PJ1401000,
and the CAS Center for Excellence in Particle Physics (CCEPP);
the Ministry of Education, Youth, and Sports of the Czech Republic under Contract No.~LTT17020 and
Charles University Grant No.~SVV 260448 and
the Czech Science Foundation Grant No.~22-18469S;
European Research Council, Seventh Framework PIEF-GA-2013-622527,
Horizon 2020 ERC-Advanced Grants No.~267104 and No.~884719,
Horizon 2020 ERC-Consolidator Grant No.~819127,
Horizon 2020 Marie Sklodowska-Curie Grant Agreement No.~700525 "NIOBE"
and
No.~101026516,
and
Horizon 2020 Marie Sklodowska-Curie RISE project JENNIFER2 Grant Agreement No.~822070 (European grants);
L'Institut National de Physique Nucl\'{e}aire et de Physique des Particules (IN2P3) du CNRS (France);
BMBF, DFG, HGF, MPG, and AvH Foundation (Germany);
Department of Atomic Energy under Project Identification No.~RTI 4002 and Department of Science and Technology (India);
Israel Science Foundation Grant No.~2476/17,
U.S.-Israel Binational Science Foundation Grant No.~2016113, and
Israel Ministry of Science Grant No.~3-16543;
Istituto Nazionale di Fisica Nucleare and the research grants BELLE2;
Japan Society for the Promotion of Science, Grant-in-Aid for Scientific Research Grants
No.~16H03968,
No.~16H03993,
No.~16H06492,
No.~16K05323,
No.~17H01133,
No.~17H05405,
No.~18K03621,
No.~18H03710,
No.~18H05226,
No.~19H00682, 
No.~22H00144,
No.~26220706,
and
No.~26400255,
the National Institute of Informatics, and Science Information NETwork 5 (SINET5), 
and
the Ministry of Education, Culture, Sports, Science, and Technology (MEXT) of Japan;  
National Research Foundation (NRF) of Korea Grants
No.~2016R1\-D1A1B\-02012900,
No.~2018R1\-A2B\-3003643,
No.~2018R1\-A6A1A\-06024970,
No.~2018R1\-D1A1B\-07047294,
No.~2019K1\-A3A7A\-09033840,
No.~2019R1\-I1A3A\-01058933,
No.~2021R1\-A4A2001897,
and
No.~2022R1\-A2C\-1003993,
Radiation Science Research Institute,
Foreign Large-size Research Facility Application Supporting project,
the Global Science Experimental Data Hub Center of the Korea Institute of Science and Technology Information
and
KREONET/GLORIAD;
Universiti Malaya RU grant, Akademi Sains Malaysia, and Ministry of Education Malaysia;
Frontiers of Science Program Contracts
No.~FOINS-296,
No.~CB-221329,
No.~CB-236394,
No.~CB-254409,
and
No.~CB-180023, and No.~SEP-CINVESTAV research Grant No.~237 (Mexico);
the Polish Ministry of Science and Higher Education and the National Science Center;
the Ministry of Science and Higher Education of the Russian Federation,
Agreement No.~14.W03.31.0026, and
the HSE University Basic Research Program, Moscow;
University of Tabuk research Grants
No.~S-0256-1438 and No.~S-0280-1439 (Saudi Arabia);
Slovenian Research Agency and research Grants
No.~J1-9124
and
No.~P1-0135;
Agencia Estatal de Investigacion, Spain
Grant No.~RYC2020-029875-I
and
Generalitat Valenciana, Spain
Grant No.~CIDEGENT/2018/020
Ministry of Science and Technology and research Grants
No.~MOST106-2112-M-002-005-MY3
and
No.~MOST107-2119-M-002-035-MY3,
and the Ministry of Education (Taiwan);
Thailand Center of Excellence in Physics;
TUBITAK ULAKBIM (Turkey);
National Research Foundation of Ukraine, project No.~2020.02/0257,
and
Ministry of Education and Science of Ukraine;
the U.S. National Science Foundation and research Grants
No.~PHY-1913789 
and
No.~PHY-2111604, 
and the U.S. Department of Energy and research Awards
No.~DE-AC06-76RLO1830, 
No.~DE-SC0007983, 
No.~DE-SC0009824, 
No.~DE-SC0009973, 
No.~DE-SC0010007, 
No.~DE-SC0010073, 
No.~DE-SC0010118, 
No.~DE-SC0010504, 
No.~DE-SC0011784, 
No.~DE-SC0012704, 
No.~DE-SC0019230, 
No.~DE-SC0021274, 
No.~DE-SC0022350; 
and
the Vietnam Academy of Science and Technology (VAST) under Grant No.~DL0000.05/21-23. 

These acknowledgements are not to be interpreted as an endorsement of any statement made
by any of our institutes, funding agencies, governments, or their representatives.

We thank the SuperKEKB team for delivering high-luminosity collisions;
the KEK cryogenics group for the efficient operation of the detector solenoid magnet;
the KEK computer group and the NII for on-site computing support and SINET6 network support;
and the raw-data centers at BNL, DESY, GridKa, IN2P3, INFN, and the University of Victoria for offsite computing support.


\bibliography{references.bib}

\clearpage

\renewcommand {\figsize} {0.7}
\newcommand {\figsuppath} {figure_supplemental}
\renewcommand{\thefigure}{S\arabic{figure}}
\setcounter{figure}{0}

\onecolumngrid

\section*{Supplementary information}

This material is submitted as supplementary information for the Electronic Physics Auxiliary Publication Service.

We provide an additional text file with numerical results of the expected background events, signal efficiency, observed yields, observed 90\% CL upper limit on the cross section of \darkh\ as well as of the observed 90\% CL upper limit on $\varepsilon^2 \times \alpha_{D}$ as functions of \aprimemass\ and \hprimemass.

\begin{figure}[htb!]
  \centering
    \includegraphics[width=\figsize\linewidth]{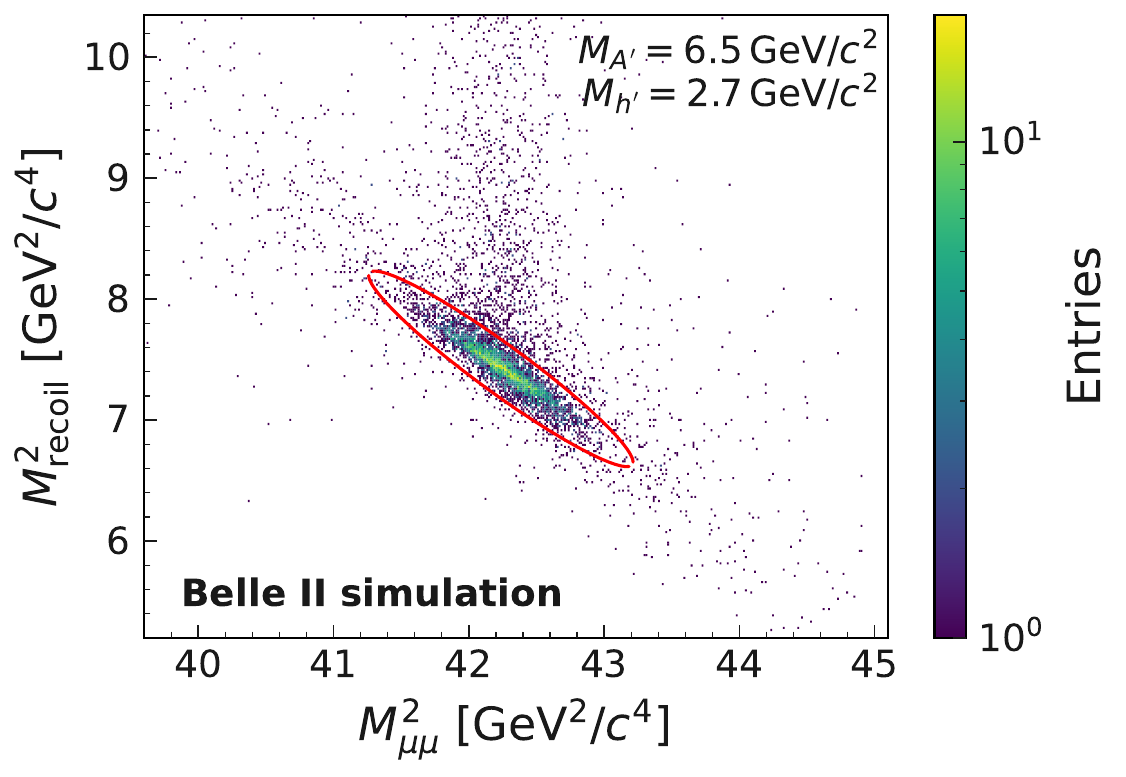}
    \caption{Example of a two-dimensional squared mass distribution for a given signal hypothesis (\aprimemass=6.5~\gevcc, \hprimemass=2.7~\gevcc). The upward tail in the squared recoil mass distribution is due to ISR. Also shown is the elliptical search window contour. }
    \label{fig:signal}
\end{figure}

\clearpage

\begin{figure}[htb!]
  \centering
    \includegraphics[width=\figsize\linewidth]{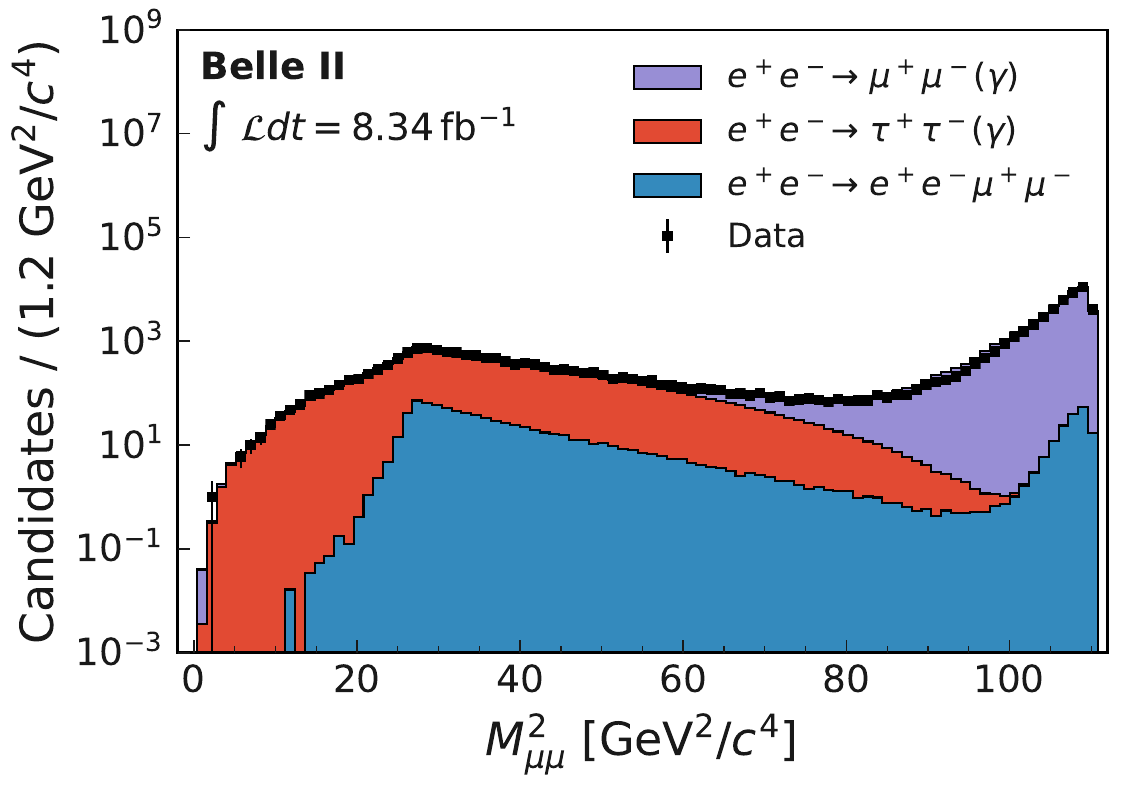}
    \caption{Squared dimuon mass $M^2_{\mu\mu}$  distribution in data and simulation, before the $C_\eta$ selection. Background contributions are stacked. }
    \label{fig:mumu_before}
\end{figure}

\begin{figure}[htb!]
  \centering
    \includegraphics[width=\figsize\linewidth]{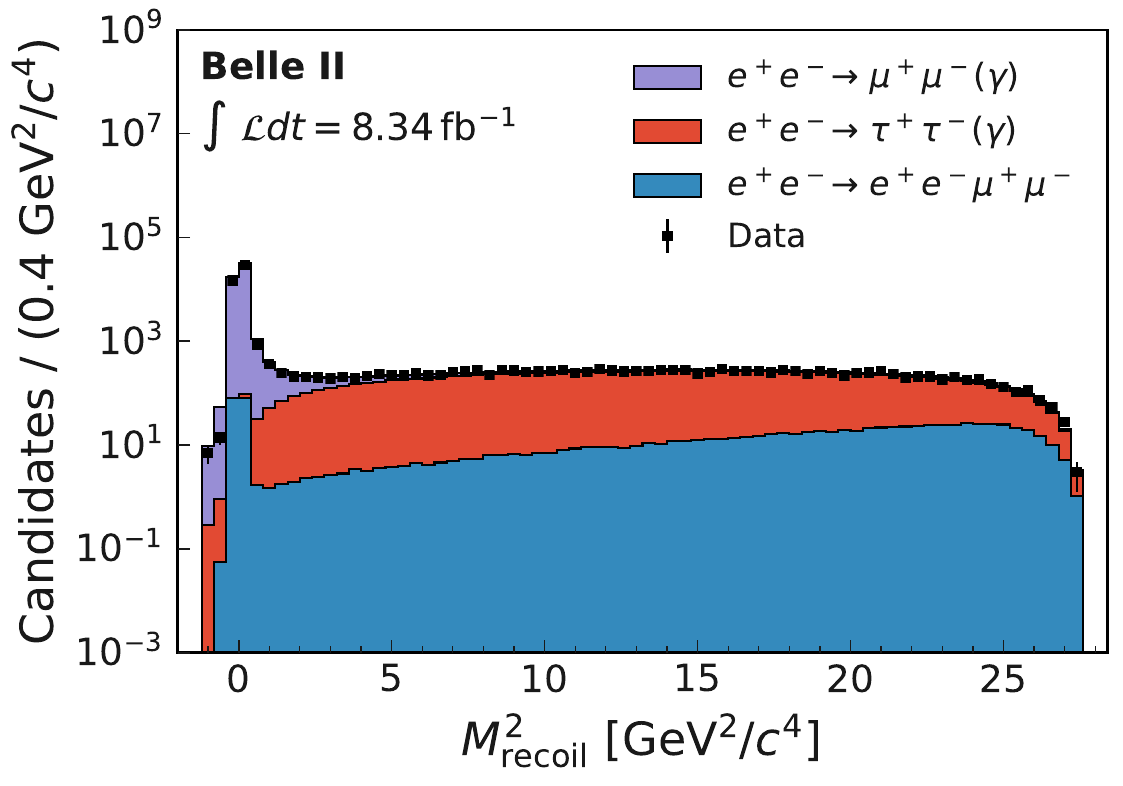}
    \caption{Squared recoil mass $M^2_{\text{recoil}}$  distribution   in data and simulation, before the $C_\eta$ selection. Background contributions are stacked.
     }
    \label{fig:mumu_before}
\end{figure}

\begin{figure}[htb!] 
  \centering
    \includegraphics[width=\figsize\linewidth]{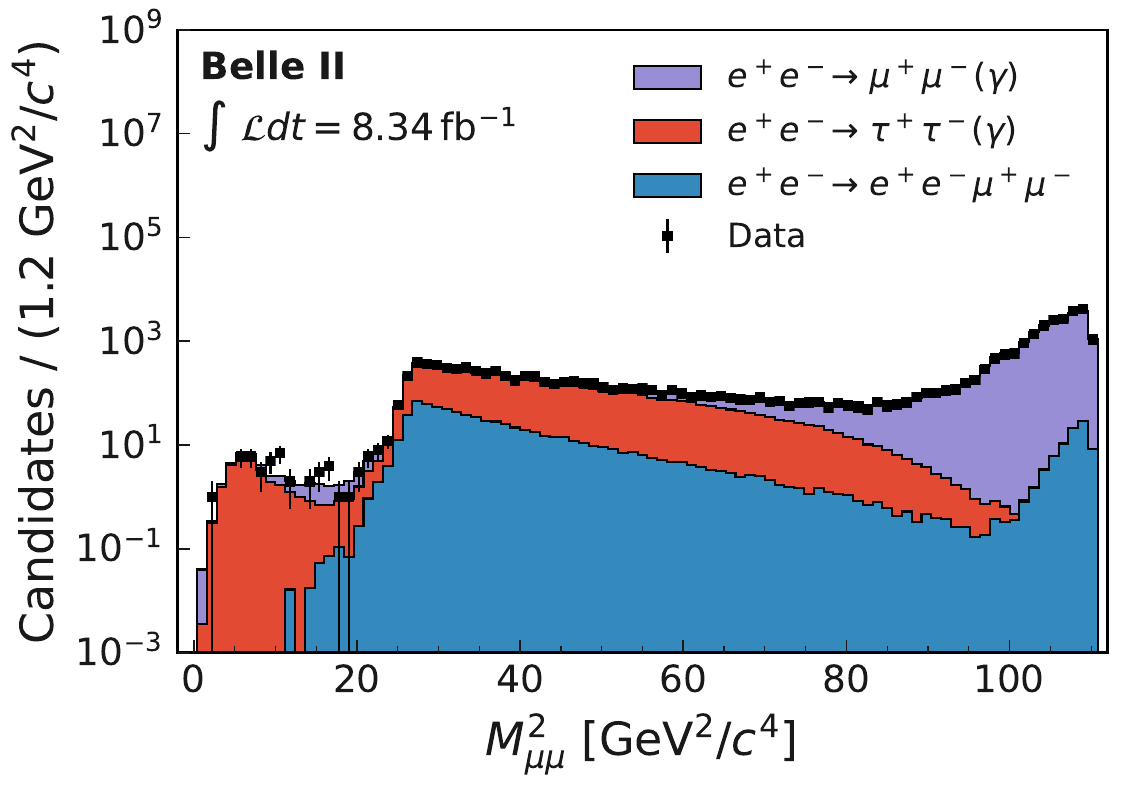}
    \caption{Squared dimuon mass $M^2_{\mu\mu}$  distribution in data and simulation, after the $C_\eta$ selection. 
    We show here events that pass the selection in at least one of the search windows in which they are contained:
    this choice is motivated by the fact that the $C_\eta$ selection is defined at search window level and windows overlap. Background contributions are stacked.}
     
    \label{fig:mumu_before}
\end{figure}

\begin{figure}[htb!]
  \centering
    \includegraphics[width=\figsize\linewidth]{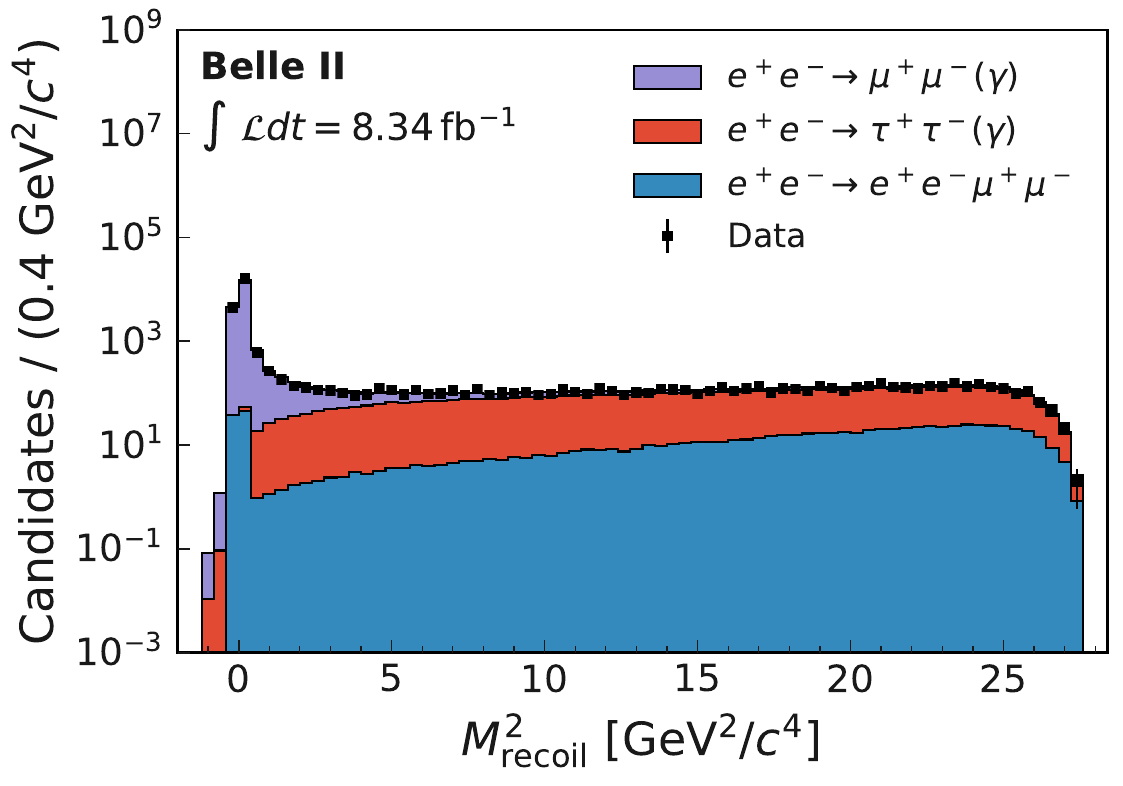}
    \caption{Squared recoil mass $M^2_{\text{recoil}}$  distribution in data and simulation, after the $C_\eta$ selection.  
    We show here events that pass the selection in at least one of the search windows in which they are contained:
    this choice is motivated by the fact that the $C_\eta$ selection is defined at search window level and windows overlap. Background contributions are stacked.
     }
    \label{fig:mumu_before}
\end{figure}

\clearpage

\begin{figure}[htb] 
  \centering
    \includegraphics[width=\figsize\linewidth]{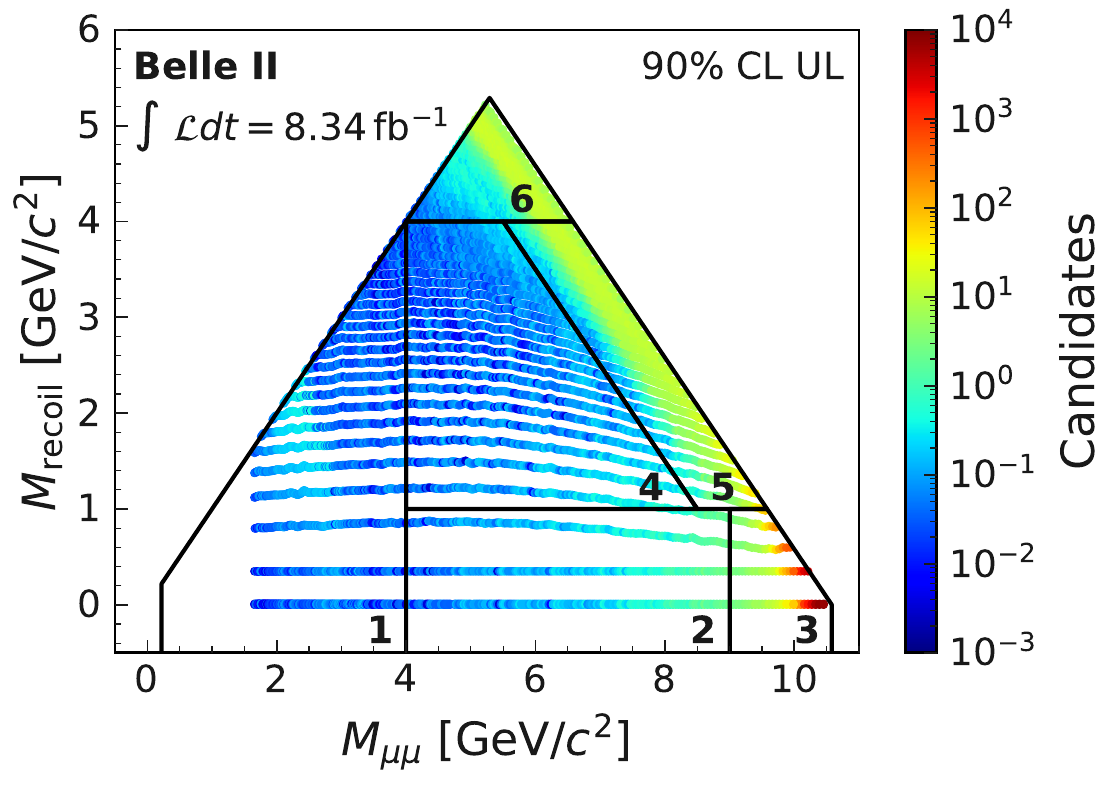}
    \caption{Expected event counts inside the search windows after all selection criteria. Points correspond to search window centers. Macroregion boundaries are also shown. } 
    \label{fig:data}
\end{figure}

\begin{table}[!htb]
  \caption{Expected and observed events in macroregions after all selection criteria. 
  We count here events that pass the selections in at least one of the search windows in which they are contained:
    this choice is motivated by the fact that the $C_\eta$ selection is defined at search window level and windows overlap.
  The dominant [subdominant, when not negligible] background source in each macroregion is also indicated. Uncertainties on expected events are from luminosity and simulation sample size.\newline}
  \centering

    \begin{tabular}{c|c|c|c}
      \hline \hline 
      Macroregion & Expected events  & Observed events  & Dominant background \\
      \hline
1 & 37.0 $\pm$ 0.7 & 35 & $\tau^+\tau^-(\gamma)$ \\
2 & 75.5 $\pm$ 1.7 & 72 & $\mu^+\mu^-(\gamma)$ \\
3 & 20779 $\pm$ 210 & 21399 & $\mu^+\mu^-(\gamma)$ \\
4 & 65.1 $\pm$ 1.4 & 71 &  $\mu^+\mu^-(\gamma)$ [$\tau^+\tau^-(\gamma)$] \\
5 & 4150 $\pm$ 42 & 4085 & $\tau^+\tau^-(\gamma)$ [$\mu^+\mu^-(\gamma)$] \\
6 & 3379 $\pm$ 34 & 3323 & $\tau^+\tau^-(\gamma)$ [$e^+e^-\mu^-\mu^-$] \\
      \hline \hline
    \end{tabular}
  \label{tab:bcks}
\end{table}

\begin{figure}[!ht]
    \centering
    \includegraphics[width=\figsize\linewidth]{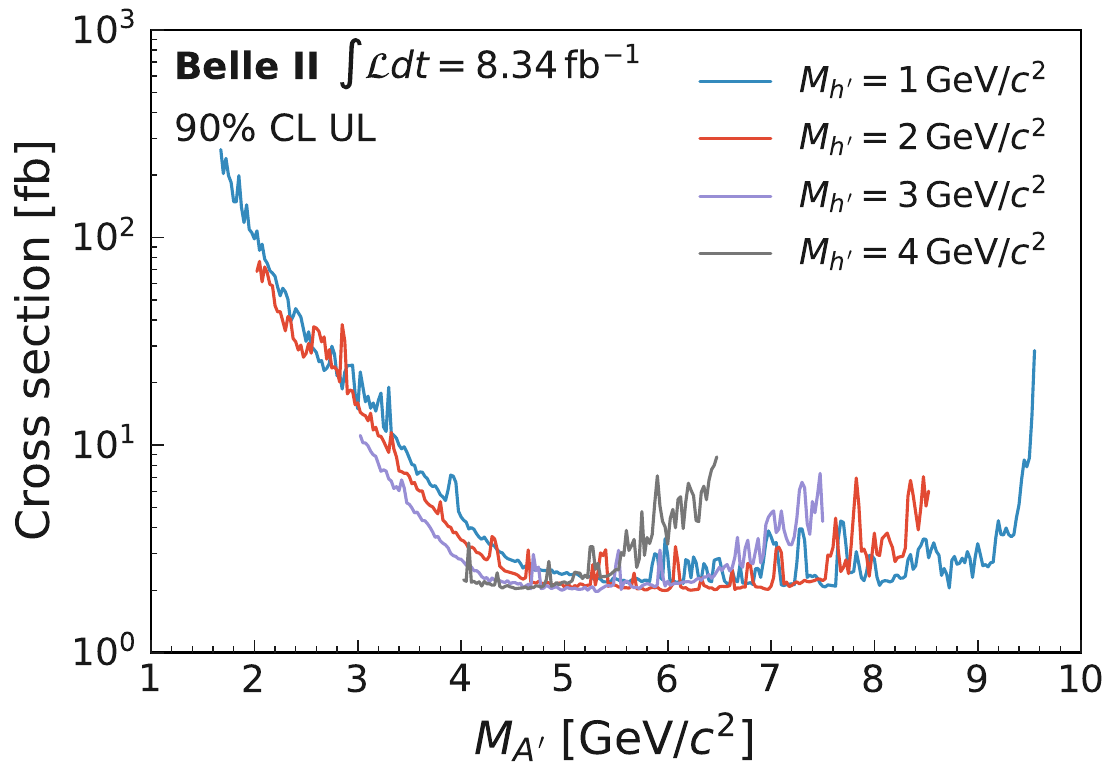} 
    \includegraphics[width=\figsize\linewidth]{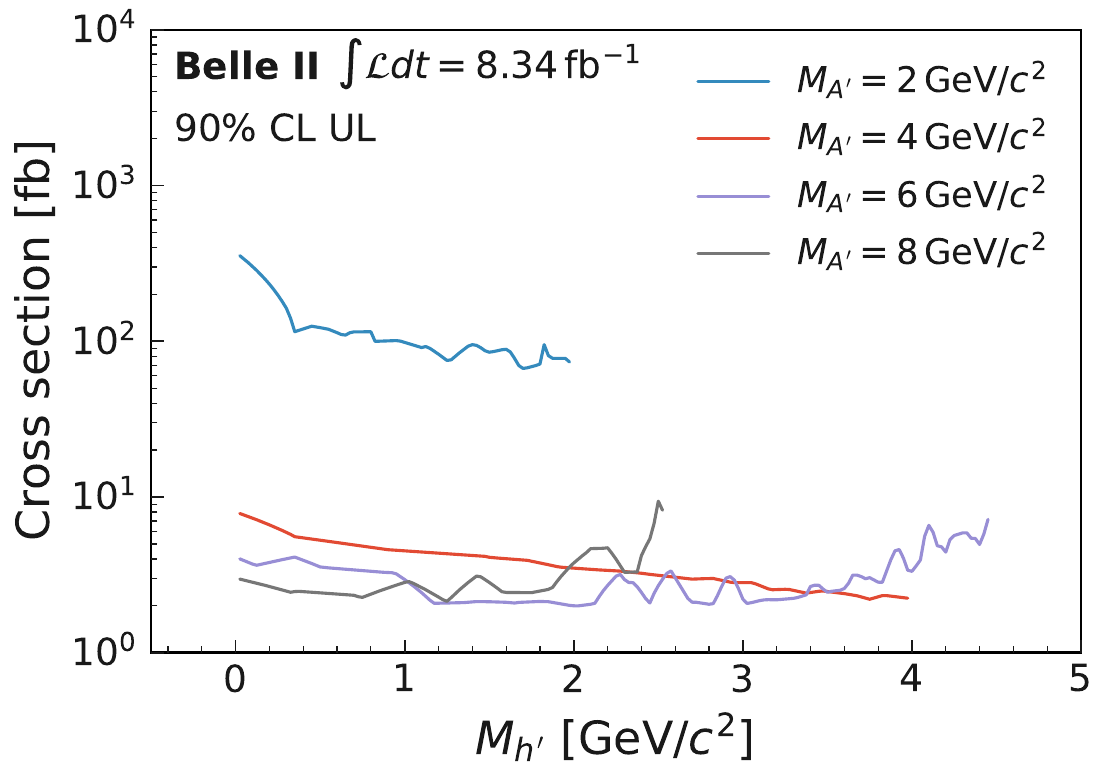}
    
    \caption{Observed 90\% CL upper limits on the cross section for \darkh\  ({\it top}) as functions of \aprimemass\ for  four values of \hprimemass\ and ({\it bottom}) as functions of \hprimemass\ for four values of \aprimemass.}
    \label{fig:xsec_projections}
\end{figure}

\begin{figure}[!ht]
    \centering
    \includegraphics[width=\figsize\linewidth]{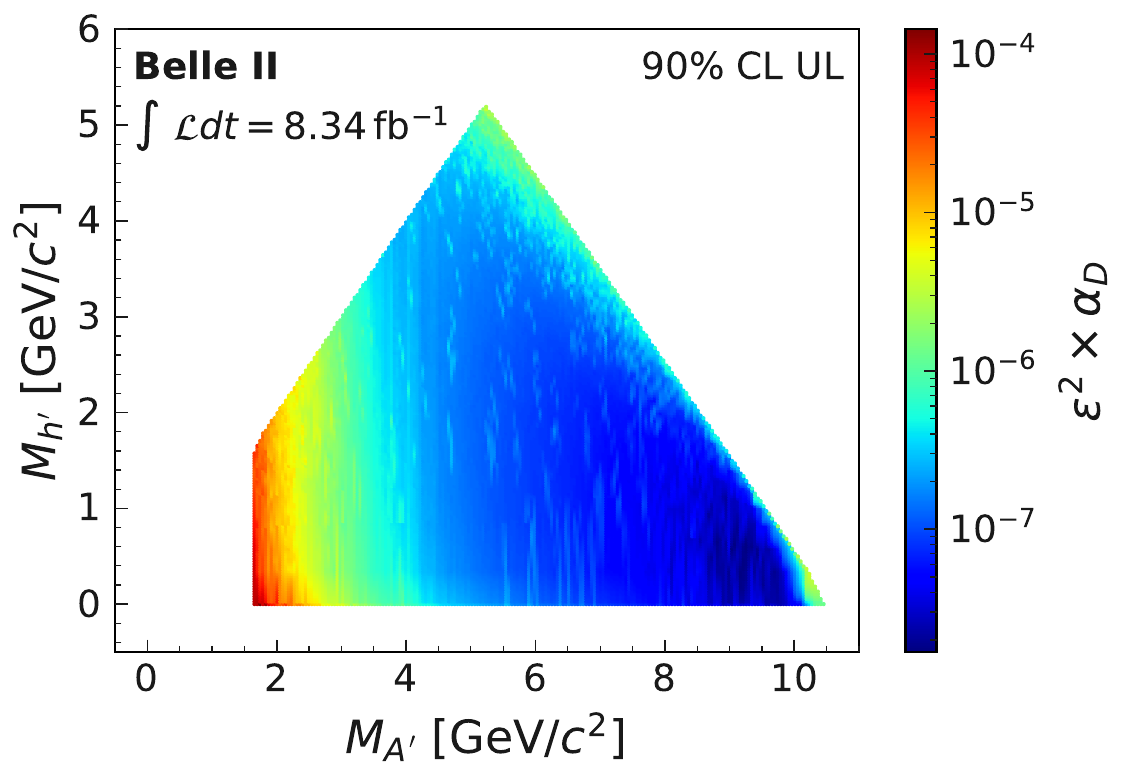}
    
    \caption{Observed 90\% CL upper limit on $\varepsilon^2 \times \alpha_{D}$ as a function of the \aprime\ and \hprime\ masses. Values are computed at search window centers and then interpolated to points of the search plane.}
    \label{fig:couplings_2d}
\end{figure}

\begin{figure}[!ht]
    \centering
    \includegraphics[width=1.01\linewidth]{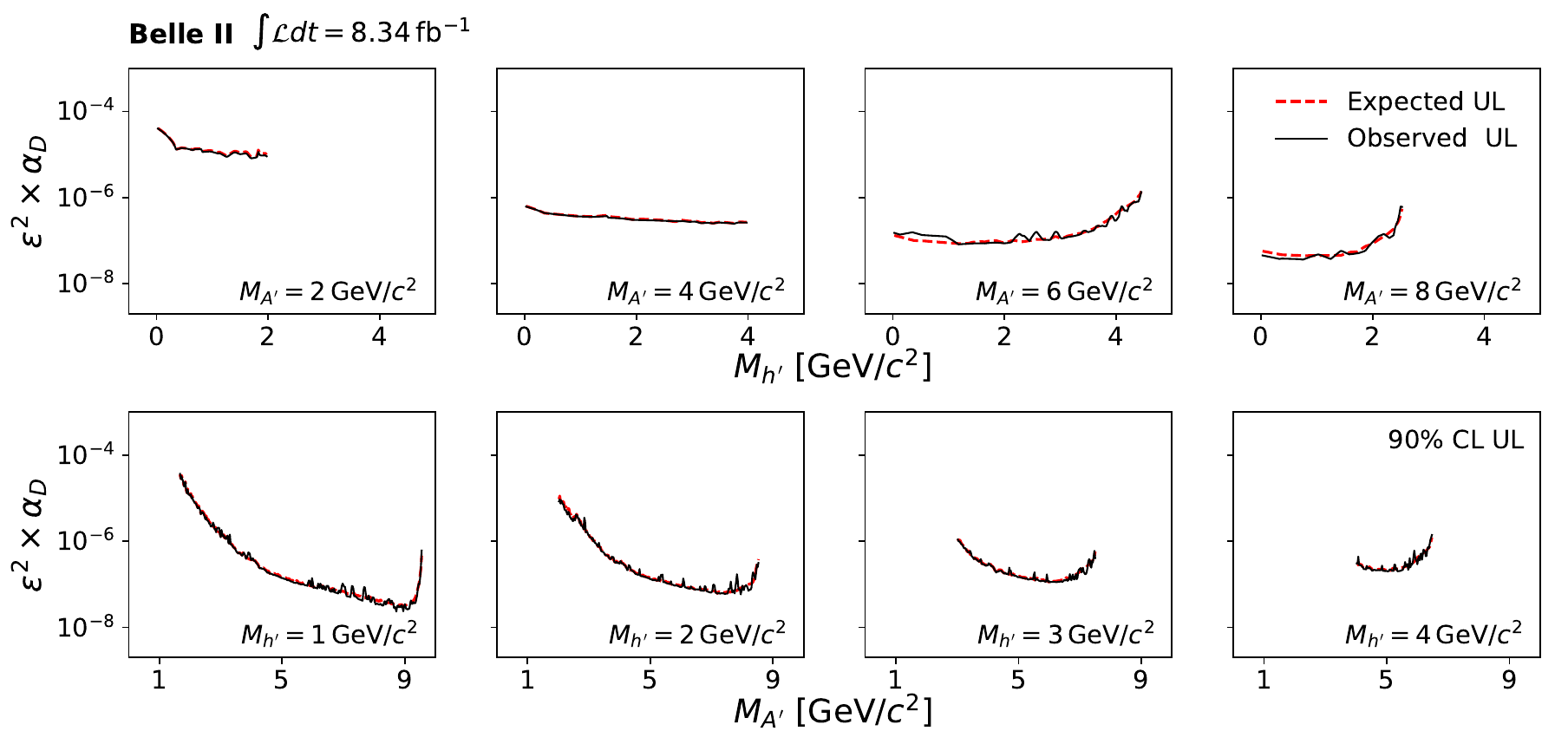}
    \caption{Observed 90\% CL upper limits on  $\varepsilon^2 \times \alpha_{D}$ (solid black line) and  their expected values (dotted red line) ({\it top}) as  functions of \hprimemass\ for four values of \aprimemass\  and  ({\it bottom}) as functions of \aprimemass\ for four values of \hprimemass.}
    \label{fig:couplings_projections_wexpected}
\end{figure}

\end{document}